\documentstyle[12pt,aasms4]{article}
\newcommand{\postscript}[2]
 {\setlength{\epsfxsize}{#2\hsize}
  \centerline{\epsfbox{#1}}}
\def\tempest%
{\begin{array}{ccc}
1 & 1 & 1 \\
1 & 1 & 1 \\
4 & 3 & 8
\end{array}}

\begin{document}

\title{The Effect of Luminous Lens Blending in \\
       Gravitational Microlensing Experiments}
\author{Seunghun Lee, Cheongho Han}
\smallskip
\affil{Dept.\ of Astronomy \& Space Science, \\
       Chungbuk National University, Cheongju, Korea 361-763 \\
       cheongho@astro.chungbuk.ac.kr,\  leess@astro.chungbuk.ac.kr}
\bigskip
\authoremail{leess@astro.chungbuk.ac.kr}
\authoremail{cheongho@astro.chungbuk.ac.kr}

\begin{abstract}
The most important uncertainty in the results of gravitational microlensing
experiments comes from the difficulties of photometry caused
by blending of source stars.
Recently Nemiroff (1997) pointed out that the results of microlensing
experiments can also be affected by the blending of light from the 
lens itself if a significant fraction of lenses are composed of stars.
In this paper, we estimate the effects of lens blending
on the optical depth determination and the derived matter distribution
toward the Galactic bulge by using
realistic models of the lens matter distribution and a well-constrained
stellar luminosity function.
We find that the effect of lens blending is largest for
lenses located in the Galactic disk.
However, lens blending does not seriously affect both the determination of
the optical depth and the Galactic matter distribution.
The decrease in optical depth is $\sim 10\%$ even under the extreme
assumption that lenses are totally composed of stars and disk matter
distribution follows a maximal disk model,
in which the lens blending effect is most severe. 
\end{abstract}

\vskip100mm
\keywords{The Galaxy --- gravitational lensing --- dark matter ---
Stars: low-mass}

\centerline{submitted to {\it The Astrophysical Journal}: July ??, 1997}
\centerline{Preprint: CNU-A\&SS-03/97}
\clearpage

\section{Introduction}
Large surveys trying to detect gravitational microlensing events
by monitoring stars located in the Large Magellanic Cloud (LMC) and
the Galactic bulge are being carried out by the MACHO (Alcock et al.\ 1997),
EROS (Renault et al.\ 1996), OGLE (Udalski et al.\ 1994), and
DUO (Alard, Mao, \& Guibert 1995) groups.\markcite{alcock1997}
\markcite{renault1996}\markcite{udalski1994}\markcite{alard1995}
They have already detected $\sim 20$ candidate events toward
the LMC and $\gtrsim 150$ events toward the Galactic bulge
(Gould, private communication).

The results of these lensing experiments are typically presented
in terms of optical depth.
By definition, the optical depth represents the average fraction
of the sources that is being gravitationally amplified more than
a factor of 1.34 at any given time.
For a given model of lens matter distribution, $\rho (D_{\rm ol})$,
the optical depth is theoretically determined by
$$
\tau = {4\pi G \over c^2}
\int_{0}^{D_{\rm os}} dD_{\rm ol} \rho (D_{\rm ol})
{D_{\rm ol}D_{\rm ls}\over D_{\rm os}},
\eqno(1.1)
$$
where $D_{\rm ol}$, $D_{\rm ls}$, and $D_{\rm os}$ are the
separations between the observer, lens, and source.
Observationally, on the other hand, the optical depth is determined by
$$
\tau =
{\pi \over 2N_\ast T}
\sum_{j=1}^{N_{\rm event}}
{t_{{\rm E},j}\over \epsilon_j},
\eqno(1.2)
$$
where $N_\ast$ and $N_{\rm event}$ are, respectively, the total numbers of
monitored stars and detected events, $T$ is the total observation time, and
$\epsilon (t_{\rm E})$ is the detection efficiency.
Here the Einstein time scale represents the required time for
a source star to cross the Einstein ring radius $r_{\rm E}$,
$$
t_{\rm E} = {r_{\rm E} \over v};\qquad
r_{\rm E} = \left( {4GM\over c^2}
{D_{\rm ol}D_{\rm ls}\over D_{\rm os}}\right)^{1/2},
\eqno(1.3)
$$
and its value is obtained by fitting theoretical light curve
to the measured  one.
The theoretical light curve is related to the lensing parameters by
$$
A_{\rm abs} = {u^2+2 \over u (u^2+4)^{1/2}};\qquad
u= \left[ \beta^2 + \left( {t-t_0 \over t_{\rm E}}\right)^2\right]^{1/2},
\eqno(1.4)
$$
where $u$ is the lens-source separation in units of $r_{\rm E}$,
$\beta$ is the impact parameter of the lens-source encounter, and
$t_0$ is the time of maximum amplification.  From the comparison of
the theoretical and observed optical depths,
one can obtain precious information about the MACHO-type matter fraction
within the Galaxy.
In addition, analysis of the optical depth distribution for
various lines of sight provides an important tool for the
study of Galactic structure (Sackett \& Gould 1993; Han \& Gould 1995).
\markcite{sackett1993}\markcite{han1995}

However, precise determination of the optical depth is hampered
by blending problem.
Due to the blended light from background stars that are not
participating in the event, the apparent amplification of
an event is lower than its absolute value.
As a result,
the apparent Einstein time scale is shorter than its true value.
Since the optical depth is directly proportional to the time scale,
the optical depth determined without proper correction of
blending effect is systematically underestimated
(Di Stefano \& Esin 1995; Wozniak \& Pacy\'nski 1997).
\markcite{stefano1995}\markcite{wozniak1997}

Recently, Nemiroff (1997)\markcite{nemiroff1997} pointed out that
not only background source stars but also bright stellar lenses
cause blending, `lens blending'.
The light from bright lenses causes the measured
optical depth to be underestimated by the same way as blending
by unresolved background stars does.
In addition, he showed that due to the lens blending the
detection of events caused by lenses closer to the observer is
comparatively more difficult than the detection of events produced
by lenses near the source.
This latter effect of lens blending causes the matter distribution
derived from the optical depth distribution to deviate from its
true value.

In this paper, we estimate the effects of lens blending
on the optical depth determination and the derived matter distribution
toward the Galactic bulge by using
realistic models of the lens matter distribution and a well-constrained
stellar luminosity function.
We find that the effect of lens blending is largest for
lenses located in the Galactic disk. 
However, lens blending does not seriously affect both the determination of
the optical depth and the Galactic matter distribution.
The decrease in optical depth is $\sim 10\%$ even under the extreme
assumption that lenses are totally composed of stars and disk matter
distribution follows a maximal disk model,
in which the lens blending effect is most severe.

\section{Blending by Bright Lenses}

If a source star is gravitationally lensed by another star, the absolute
amplification, $A_{\rm abs}$, of the event is diluted by the light
from the lens, and the event appears to be amplified by
$$
A_{\rm app} =
{A_{\rm abs}l_S + l_L \over l_S + l_L },
\eqno(2.1)
$$
where $l_L$ and $l_S$ are the apparent fluxes of the lens
and the source and $A_{\rm abs}$ is the absolute amplification.
Due to this dilution of amplification, the measured event time scale
appears to be shorter than its true value by a factor
$$
\eta =
\left[ 2\left( 1-A_{\rm min}^{-2} \right)^{-1/2}-2 \right]^{1/2};\qquad
A_{\rm min} = 1.34 \left( 1 + {l_L\over l_S} \right) -
\left( {l_L\over l_S} \right),
\eqno(2.2)
$$
where $A_{\rm min}$ is the minimum required amplification
for detection (Nemiroff 1997)\markcite{nemiroff1997}.
Since the optical depth is directly proportional to
the time scale, $\tau$ is decreased by the same factor by the lens blending.

In addition, lens blending makes
the distribution of lenses causing events along the line of sight
toward a direction differ from its true one.
In general, the lens matter distribution does not coincide with
the matter distribution.
This difference between the lens and matter distributions arises
because the cross-section of the lens-source encounter,
i.e., Einstein ring diameter $2r_{\rm E}$,
depends on the geometry of the lens system by equation (1.3).
For a uniform distribution of matter, the lensing probability peaks at
the very center between the observer and the source star, and it decreases
as the lens approaches either the source or the lens.
In addition to this intrinsic dependence on the lens system geometry,
the optical depth distribution additionally depends on the lens 
location due to the effect of lens blending.
This additional dependency on lens geometry arises because
stellar lenses that are nearby, and thus apparently brighter,
cause more blending than those near the sources do
(Nemiroff 1997).\markcite{nemiroff1997}
Therefore, the factor by which the optical depth is decreased depends
not only on the relative absolute lens/source flux ratio,
$L_L /L_S$, but also on the lens geometry, i.e.,
$$
\eta \left( {L_L\over L_S},{D_{\rm os}\over D_{\rm ol}}\right) =
\left[ 2\left( 1-{A'}_{\rm min}^{-2} \right)^{-1/2}-2 \right]^{1/2};\qquad
{A'}_{\rm min} = 1.34 \left[ 1+ {L_L\over L_S}
\left( {D_{\rm os}\over D_{\rm ol}}\right)^2 \right] -
{L_L\over L_S}
\left( {D_{\rm os}\over D_{\rm ol}}\right)^2.
\eqno(2.3)
$$

\section{Models of Matter Distribution and Luminosity Function}

Since the effect of lens blending depends on both the
lens geometry and the lens/source flux ratio,
it is required to model the Galactic matter distribution
and the luminosity function of stars for the estimation of
the lens blending effect on actual lensing experiments.
In addition, the fraction of stellar lenses
among all types (dark+luminous) of lenses should be known.
However, all these quantities required to estimate
the lens blending effect are poorly known.
Therefore, we test various models of matter distribution, lens brightness,
and stellar lens fraction.

A fraction of events toward the LMC might be caused by stars in the
Galactic disk and in the LMC itself (Sahu 1994).\markcite{sahu1994}
However, this fraction of events is $\lesssim 10\%$ of the total events 
expected from all-MACHO halo (Wu 1994).\markcite{wu1994}
On the other hand, a significant fraction of events toward
the Galactic bulge lenses is thought to be caused by stars
(Kamionkowski 1995; Han, Chang, \& Lee 1997).
\markcite{kamionkowski1995}\markcite{han1997b}
Therefore, we estimate the effect of the lens blending only on the
experiments which are being carried out toward the Galactic bulge.

We test two bulge matter distribution models.
First, we adopt the axisymmetric Kent bulge model
(Kent 1992)\markcite{kent1992} of the form,
$$
\rho (s)=
\cases{
1.04\times 10^6 (s/0.482)^{-1.85}\ M_\odot\ {\rm pc}^{-3}, &
 (for $s < 938\ {\rm pc}$), \cr
3.53K_0\ (s/667)\ M_\odot\ {\rm pc}^{-3}, &
(for $s \geq 938\ {\rm pc}$), \cr
}
\eqno(3.1)
$$
where $s^4=R^4+(z/0.61)^4$, $R=(x^2+y^2)^{1/2}$, and the axes $x$ and $z$
are directed toward the observer and toward the Galactic pole from the
Galactic center, respectively.
The second COBE bulge model (Dwek et al.\ 1995)\markcite{dwek1995}
has a triaxial shape, or bar-shaped, with an analytic form of
$$
\rho (r_s)= \rho_{0,{\rm COBE}}  \exp (0.5r_s^2)\ M_\odot\ {\rm pc}^{-3},
\eqno(3.2)
$$
where $r_s= \lbrace [(x'/x_0)^2+(y'/y_0)^2]^2+(z'/z_0)^4\rbrace^{1/4}$ and
$(x_0,y_0,z_0)=(1.58, 0.62, 0.43)\ {\rm kpc}$.
The coordinates $(x',y',z')$ represent axes of the bar from the
longest to the shortest, and the longest axis is misaligned with the $x$ axis
by an angle of $20^\circ$.
We set the normalization to be
$\rho_{\rm COBE} \sim 2.0\times 10^9\ M_\odot\ {\rm pc}^{-3}$ so that the
total mass of the bulge matches with the Zhao et al.'s (1995)\markcite{zhao}
estimate of $2.0\times 10^{10}\ M_\odot$.

For the Galactic disk matter distribution, we adopt a double exponential
disk of the form
$$
\rho (R,z) = 0.06\exp\left\{
-\left[ {R-8000\over h_R}+{z\over h_z}\right]\right\}
\ M_\odot\ {\rm pc}^{-3},
\eqno(3.3)
$$
where the radial and vertical scale heights are
$h_R=3.5\ {\rm kpc}$ and $h_z=325\ {\rm pc}$
(Bahcall 1986).\markcite{bahcall1986}
Since the events caused by stars in the disk are more likely to be
affected by the lens blending effect compared to those caused by
relatively remote bulge stars, we also test the maximal disk model.
In the maximal disk model, matter is distributed in the same way as
in the Bahcall disk model, but the distribution has 3 times higher
normalization.
The adopted matter distribution models of the Galactic structures 
are summarized in Table 1.

By using these models, we test  3 combinations of Galactic matter 
distribution.
These combinations are Kent bulge + Bahcall disk (model I),
COBE bulge + Bahcall disk (model II), and Kent 
bulge + maximal disk (model III) as listed in Table 2.
The matter distributions along the line of sight toward
Baade's Window located at $(\ell ,b)=(1^\circ, -3^\circ\hskip-2pt .9)$
for these three Galactic mass distribution models are shown in 
Figure 1.
One finds that model II has lower disk/bulge matter ratio 
compared to model I distribution, while model III has higher ratio.

The $I$--band luminosity function of stars is constructed by combining
ground (J.\ Frogel, private communication) and
space-based observations (R. M. Light, private communication).
In addition, we extend the very faint part of the luminosity
function by adopting that of stars in the solar neighborhood
(Gould, Bahcall, \& Flynn 1996).\markcite{gould1996}
For the construction of the luminosity function, we assume that
the populations of stars in the disk and bulge are similar each other.
The finally constructed luminosity function is presented in Fig 1 of
Han (1997).\markcite{han1997a}

\section{Estimate of Lens Blending Effect}

Based on the models of Galactic mass distribution and the luminosity function,
the optical depth distribution taking lens blending into consideration
is computed by
$$
{d\tau (D_{\rm ol})\over d D_{\rm ol}}  =
{4\pi G\over c^2} \rho (D_{\rm ol}) 
\int_{D_{\rm ol}}^{d_{\rm max}}
dD_{\rm os} \eta \left( {L_L\over L_S},
{D_{\rm os}\over D_{\rm ol}}\right)
n(D_{\rm os}){D_{\rm ol}D_{\rm ls}\over D_{\rm os}}
\times  \left[ \int_{D_{\rm ol}}^{d_{\rm max}}
dD_{\rm os} n(D_{\rm os})\right]^{-1},
\eqno(4.1)
$$
where $n(D_{\rm os})\propto \rho (D_{\rm os})$ is the number density of 
source stars and
$d_{\rm max}=12\ {\rm kpc}$ is the upper limit of the source
star distribution.
Above equation differs from equation (1.1) because source stars are
distributed in a wide spatial range, and thus their locations can
no longer be approximated as a constant.
The term in the bracket ([ ]) is included to normalize
the lensing probability for a single source star.
In addition, by setting the lens mass density to be the sum
of disk and bulge densities, i.e.,
$\rho = \rho_{\rm disk} + \rho_{\rm bulge}$,
we include both disk-bulge and self-lensing (disk-disk and bulge-bulge)
event contribution in the optical depth computation.

Recently Han et al.\ (1997) determined the fraction of Galactic bulge
events caused by stars to be $f_\ast\lesssim 50\%$.
Therefore, in our computation we assume $50\%$ of events are
due to stellar lenses, and the other 50\% of events are
caused by dark component of lenses.
However, the luminous lens fraction is still very uncertain.
Therefore, we also test two other cases of lens fractions in which
stars comprise 70\% and 100\% of total lenses.
In the computation, we choose the brightnesses of lenses
from the model luminosity function by a Monte Carlo method.
On the other hand, source star brightnesses are selected from the 
part of the luminosity function brighter than
the $I$--band absolute magnitude of $M_I=4$, which is the
detection limit of ground based observation. 
This is because while stars at any brightness can work as lenses,
to be monitored source stars should be
bright enough to be resolved overcoming the threshold detection
limit imposed by the blending of stars.

The optical depth distributions for individual mass distribution models
are computed by equations (2.3) and (4.1), and they are shown in
the upper panels of Figure 2.
In each panel, we compute the optical depth distributions with
the stellar lens fractions of $f_\ast = 50\%$ (dotted line),
70\% (short-dashed line), and 100\% (long-dashed line),
and they are compared to the distribution obtained without any
lens blending effect, i.e., $f_\ast = 0\%$, (solid line).
To better show the slight differences in the optical depth distributions
for different values of stellar lens fraction, the
sections of distributions in the range
$2\ {\rm kpc} \leq D_{\rm ol}\leq 4\ {\rm kpc}$ are magnified and
presented in the middle panels.
In addition, the total optical depths for individual mass distribution
models with (and without) the lens blending effect are computed by
$$
\tau_{\rm with(w/o)}=
\int_0^{d_{\rm max}} {\tau (D_{\rm ol})\over dD_{\rm ol}} dD_{\rm ol},
\eqno(4.2)
$$
and they are listed in Table 3.

One finds that the blending effect on optical depth distribution mainly
occurs for disk lenses as pointed out by Nemiroff (1997).\markcite{nemiroff1997}
This can be easily seen in the optical depth distribution ratio,
$d\tau_{\rm with} (D_{\rm ol})/d\tau_{\rm w/o} (D_{\rm ol})$, shown
in the lower panels of Figure 2.
As the disk/bulge mass ratio increases, lensing events are more likely
to be affected by lens blending, resulting in smaller values of optical depth.
For example, the ratio between the total optical depths with and without
considering the blending effect for the model III mass distribution
is smaller than that of model I 
However, for a fixed stellar lens fraction the decreases in optical depth
for different mass distribution models are similar one another regardless
of the assumed models.
That is, lens blending effect is fairly mass-distribution independent.
On the other hand, lens blending effect has relatively strong dependency
on the assumed stellar lens fractions.
Under model I mass distribution, for example, the optical depth ratio is
$\tau_{\rm with}/\tau_{\rm w/o} = 94.3\%$ if half of lenses are composed
of stars, i.e., $f_\ast = 50\%$, while
$\tau_{\rm with}/\tau_{\rm w/o} = 88.8\%$ for pure stellar lenses,
i.e., $f_\ast =100\%$.

However, the result of lensing experiments is not seriously affected
by the lens blending.
The decrease in optical depth is just $\sim 5\%$ for the most probable 
stellar lens fraction of $f_\ast = 50\%$.
Even for the extreme case where the lenses are totally composed of stars
and the disk matter follows the maximal distribution, in which the lens
blending has maximum effect, the decrease in $\tau$ is $\sim 13.5\%$.
The reason for this small effect of lens blending is that
although a significant fraction of lenses can be composed
of stars, most of them are very faint.
On the other hand, the source stars being monitored by current
lensing experiments are relatively very bright compared to typical
stellar lenses.
Therefore, the approximation of dark lenses is still good enough
for the current experiments' purpose of determining dark matter fraction
and studying Galactic structures.

\acknowledgements
We would like to thank S. Gaudi \& A. Berlind for making precious comments and
suggestions.

\clearpage

\begin{center}
\bigskip
\bigskip
\centerline{\small {TABLE 1}}
\smallskip
\centerline{\small {\sc The Matter Density Distribution Models}}
\smallskip
\begin{tabular}{ll}
\hline
\hline
\multicolumn{1}{c}{model} &
\multicolumn{1}{c}{distribution} \\
\hline
{\bf bulge} &  \\
Kent     & $\rho (s) = 1.04\times 10^6 (s/0.482)^{-1.85}
            \ M_\odot\ {\rm pc}^{-3}
            \ \ \ ({\rm for}\ s<938\ {\rm pc})$ \\
         &  $\rho (s) = 3.53K_0 (s/667)\ M_\odot\ {\rm pc}^{-3}
             \ \ \ \ \ \ \ \ \ \ \ \ \ \ \  ({\rm for}\ s\geq 938\ {\rm pc})$ \\
\bigskip
COBE     & $\rho (r_s) = 2.0\times 10^9\exp (0.5 r_s^2)\
                M_\odot\ {\rm pc}^{-3}$ \\ {\bf disk} & \\
                Bahcall  & $\rho (R,z) = 0.06
                \exp\lbrace -[(R-8000)/3500 + z/325]\rbrace
                \ M_\odot\ {\rm pc}^{-3}$ \\
Maximal  & $\rho (R,z) = 0.18
                 \exp\lbrace -[(R-8000)/3500 + z/325]\rbrace
                \ M_\odot\ {\rm pc}^{-3}$ \\
\hline
\end{tabular}
\end{center}
\smallskip
\noindent
{\footnotesize
\qquad NOTE.---
The model Galactic bulge and disk matter distributions.
In the Kent bulge model, $s^4 = R^4+(z/0.61)^4$, $R=(x^2+y^2)^{1/2}$,
and $x$ and $z$ axes direct toward the observer and toward the
Galactic pole.
In the COBE bulge model,
$r_s = \lbrace [(x'/x_0)^2 + (y'/y_0)^2]^2 +(z'/z_0)^4\rbrace^{1/4}$,
where $(x_0,y_0,z_0)=(1.58,0.62,0.43)\ {\rm kpc}$, and
$x'$ and $z'$ represent the longest and shortest axes
of the triaxial bulge.
For both bulge models, the normalizations are set so that the total mass
of the bulge is $M_{\rm bulge}\sim 2\times 10^{10}\ M_{\odot}$.
In the maximal disk model, matter is distributed the same way as
in the Bahcall disk model, but has 3 times higher normalization.
}

\newpage
\begin{center}
\bigskip
\bigskip
\centerline{\small {TABLE 2}}
\smallskip
\centerline{\small {\sc The Galactic Mass Distribution Model}}
\smallskip
\begin{tabular}{cccc}
\hline
\hline
\multicolumn{1}{c}{model} &
\multicolumn{2}{c}{component} \\
\multicolumn{1}{c}{ } &
\multicolumn{1}{c}{bulge} &
\multicolumn{1}{c}{disk} \\
\hline
I & Kent & Bahcall \\
II & COBE & Bahcall \\
III & Kent & Maximal \\
\hline
\end{tabular}
\end{center}
\bigskip
\noindent
{\footnotesize
\qquad NOTE.---
The Galactic mass distribution models.
They are obtained from the combination of the Galactic bulge and disk 
models listed in Table 1.
}

\newpage
\begin{center}
\bigskip
\bigskip
\centerline{\small {TABLE 3}}
\smallskip
\centerline{\small {\sc The Effect of Lens Blending on Optical Depth 
Determination}}
\smallskip
\begin{tabular}{ccccc}
\hline
\hline
\multicolumn{1}{c}{$f_\ast$} &
\multicolumn{1}{c}{mass} &
\multicolumn{1}{c}{$\tau_{\rm with}$} &
\multicolumn{1}{c}{$\tau_{\rm w/o}$} &
\multicolumn{1}{c}{$\tau_{\rm with}/\tau_{\rm w/o}$} \\
\multicolumn{1}{c}{($\%$)} &
\multicolumn{1}{c}{model} &
\multicolumn{1}{c}{($10^{-6}$)} &
\multicolumn{1}{c}{($10^{-6}$)} &
\multicolumn{1}{c}{($\%$)} \\
\hline
50\%  & I   & 1.018 & 1.079 & 94.3 \\
      & II  & 1.334 & 1.411 & 94.5 \\
\bigskip
      & III & 1.684 & 1.808 & 93.1 \\

70\%  & I   & 0.994 & 1.079 & 92.1 \\
      & II  & 1.304 & 1.411 & 92.4 \\
\bigskip
      & III & 1.636 & 1.808 & 90.5 \\

100\% & I   & 0.958 & 1.079 & 88.8 \\
      & II  & 1.259 & 1.411 & 89.2 \\
      & III & 1.564 & 1.808 & 86.5 \\
\hline
\end{tabular}
\end{center}
\bigskip
\noindent
{\footnotesize
\qquad NOTE.---
The effect of lens blending on  the determination of the optical depth.
Here $f_{\ast}$ represents the stellar lens fraction among all 
(luminous+dark) types of lenses.
Therefore, the stellar lens fraction of 100\% represents
that lenses are totally composed of pure stars.
}

\bigskip
\postscript{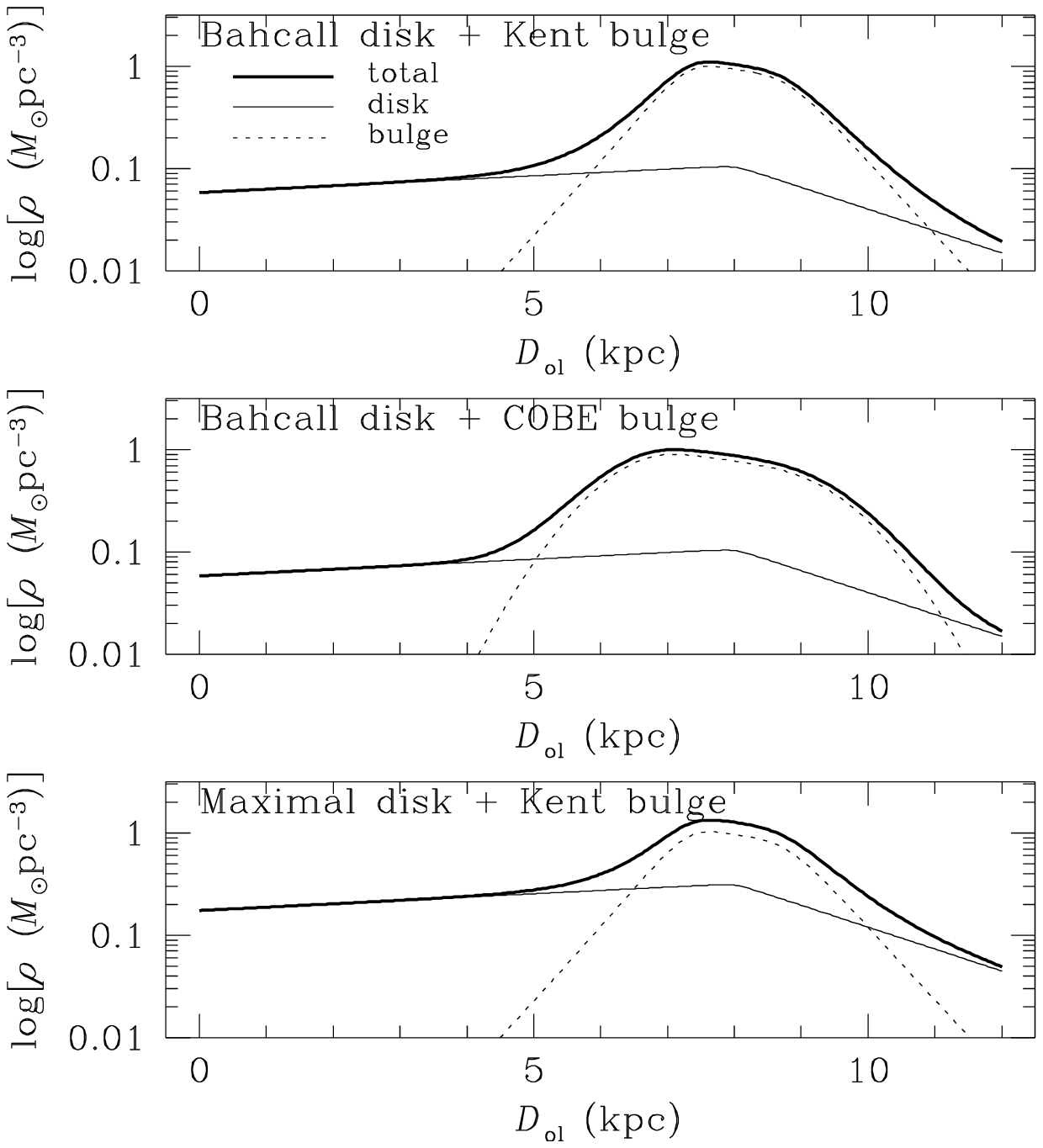}{0.72}
\noindent
{\footnotesize {\bf Figure 1:}\
The mass distributions along the line of sight toward Baade's Window for
various Galactic mass distribution models listed in Table 2.
The distributions of the bulge and disk matter are represented, respectively,
by thin dotted and solid lines, and the thick solid represents the sum 
of both distributions.
}

\clearpage

\postscript{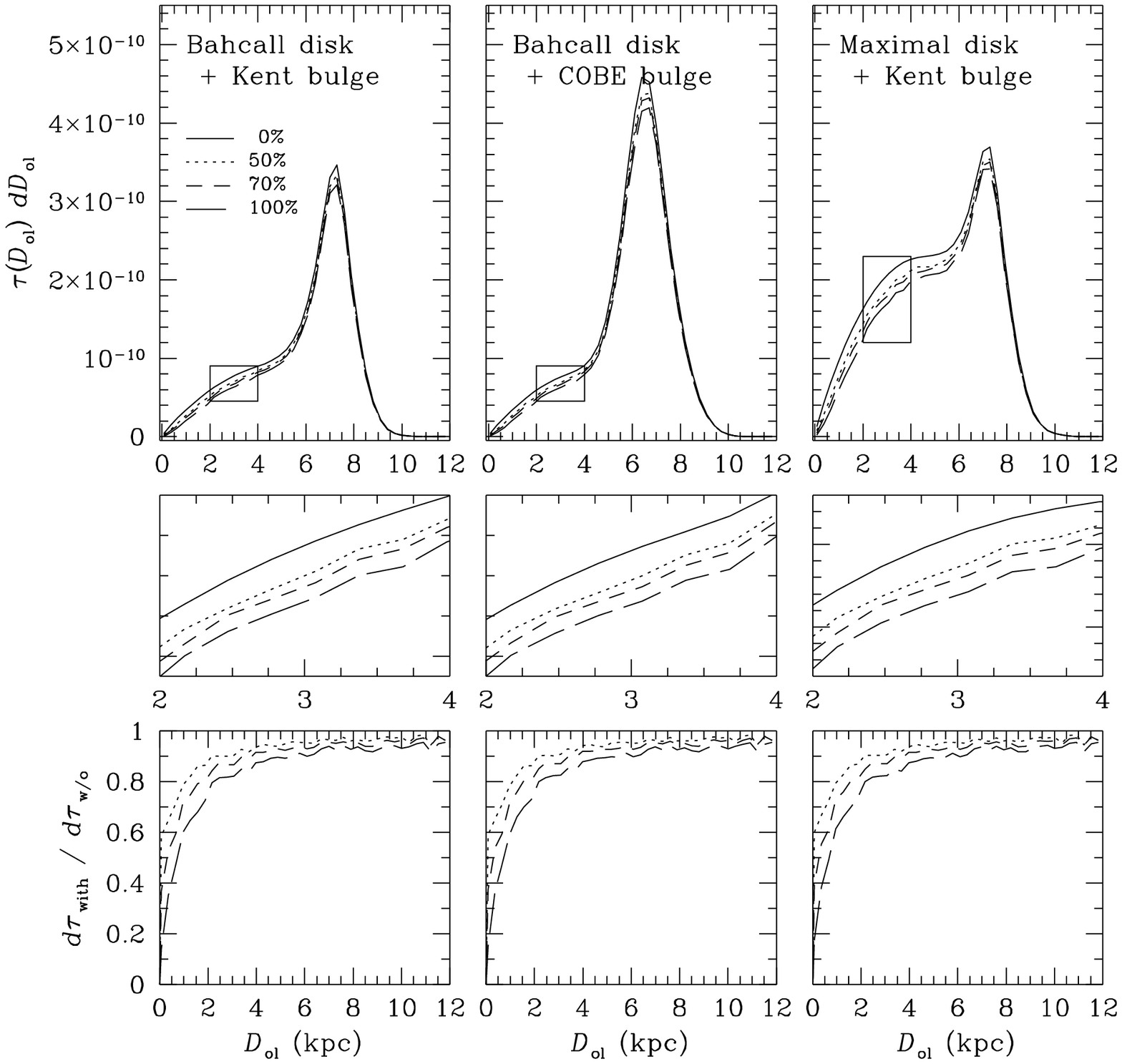}{0.81}
\noindent
{\footnotesize {\bf Figure 2:}\
The optical depth distributions for various models of Galactic 
mass distribution and stellar lens fractions.
To better show the slight differences in the optical depth distribution,
the sections of small boxes in the individual upper panels are 
magnified and shown in corresponding middle panels.
Also shown in the lower panels are the ratios between the optical 
depth distribution with and without the blending effect.
}

\end{document}